\documentclass[preprint]{vgtc}               %

\graphicspath{{figures/}{pictures/}{images/}{./}} %

\usepackage{times}                     %

\usepackage{tabu}                      %
\usepackage{booktabs}                  %
\usepackage{lipsum}                    %
\usepackage{mwe}                       %
\usepackage{amsmath}

\usepackage{mathptmx}                  %

\onlineid{0}

\vgtccategory{Research}

\vgtcinsertpkg

\title{VisPubs Games: Joyful Discovery of Visualization Research(ers)
}

\author{Devin Lange\thanks{e-mail: devin@hms.harvard.edu}\\ %
        \scriptsize Harvard Medical School %
\and Zach Cutler\thanks{e-mail: zcutler@sci.utah.edu}\\ %
      \scriptsize University of Utah %
\and Maxim Lisnic\thanks{e-mail: mlisnic@wpi.edu}\\ %
      \scriptsize Worcester Polytechnic Institute %
}

\teaser{
  \centering
  \includegraphics[width=\linewidth]{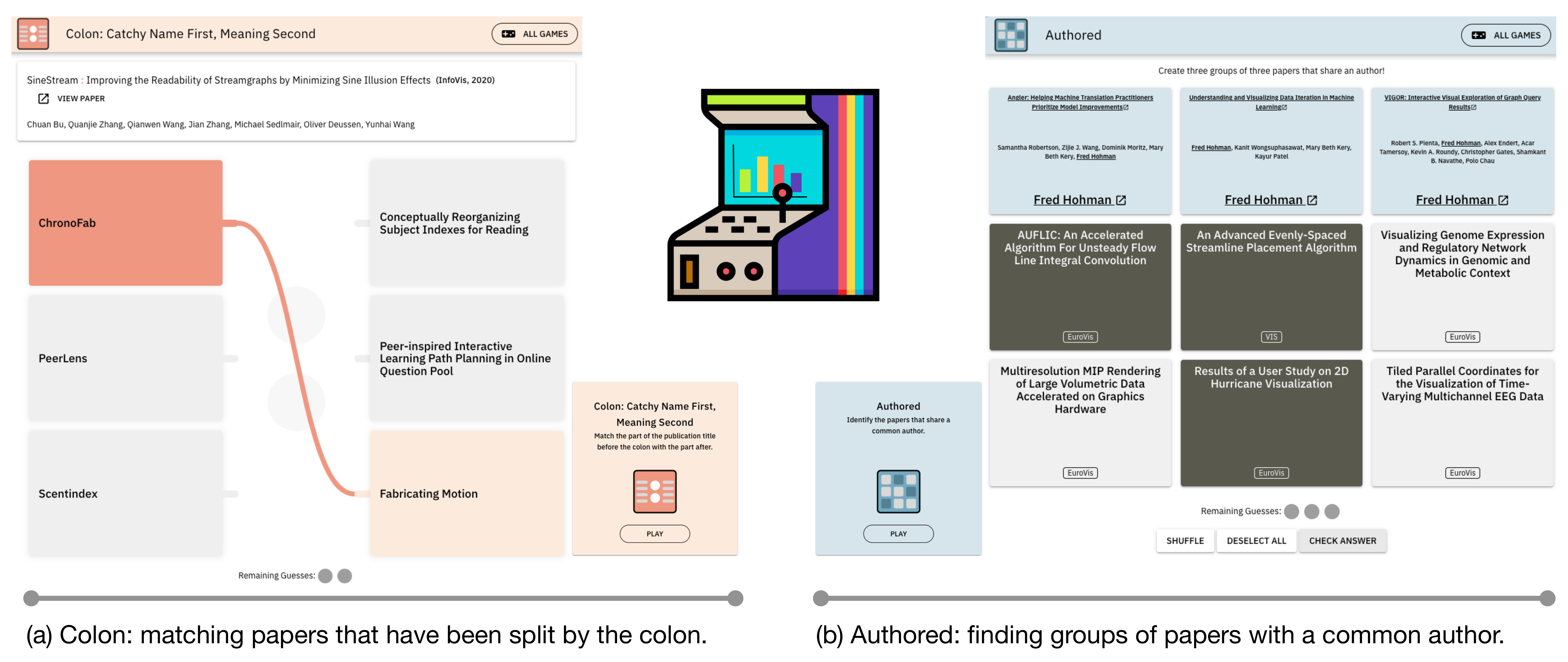}
  \caption{Screenshots from the two games we developed, Colon and Authored.}
  \label{fig:teaser}
}

\abstract{
    Many sophisticated tools exist to help researchers find the academic literature they are searching for, but what about finding work that you aren't looking for?
We promote joyful discovery of visualization research through two games (Colon and Authored) available to play now at https://games.vispubs.com.
We believe these games provide several benefits to the visualization research community.
First, the joyful discovery of visualization research and researchers occurs because these games randomly select authors and publications, thus exposing players to research areas they may not typically engage with.
Second, these games were made by visualization researchers for visualization researchers --- playing this game, sharing results with friends in person and online, has the potential to strengthen our academic community.
Third, games centered around publication authors provide a passive way for academics to gain exposure within the community.
Finally, we hope these games are simply fun to play.
Try them now at \href{https://games.vispubs.com}{games.vispubs.com}.

} %


\begin{document}
\firstsection{Introduction}
\maketitle

Long story short, we created two games --- Colon and Authored --- which are related to visualization publications. The games are available to play at \href{https://games.vispubs.com}{games.vispubs.com}. This paper describes our motivation for building these games, the game designs, and our reflections on playing them.

As data visualization researchers, we have all experienced the task of literature review in our community. This task has already inspired various academic works. Our community has devoted energy to curating datasets of visualization publications \cite{isenberg_vispubdata.org_2017, lange_vispubs.com} and building visualization tools and interfaces for exploring the landscape of research publications \cite{ponsard_paperquest_2016, beck_puresuggest_2025, lange_vispubs.com, beck_visual_2016, narechania_vitality_2022, lee_paperweaver_2024}. Furthermore, these resources have been utilized to understand the field and community of visualization research \cite{hao_thirty-two_2023, sarvghad_scientometric_2023, isenberg_visualization_2017}.

We take an \textbf{alt}ernative approach to interacting with this type of visualization publication metadata --- we built a couple of games. Despite joy and fun often being associated with ``unseriousness'' we believe that these games will provide several positive contributions to the visualization research community, and their playful nature makes them more beneficial, not less. We are not alone in utilizing the strengths of games. For instance, VisFutures \cite{dalton_visfutures_2023} is a card game that explores the future of data visualization.

\begin{figure}[t]
	\centering
	\includegraphics[width=\linewidth]{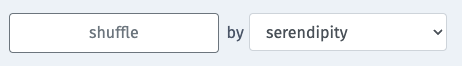}
	\caption{The default sorting option on the IEEE VIS website is serendipitous --- it shuffles the paper order randomly.}
	\label{fig:vis_shuffle_serendipity}
\end{figure}

\textbf{Joyful discovery} of visualization research or researchers is the most academic-leaning contribution of this work. A common approach for finding papers involves searching for matching keywords in paper titles or abstracts, or filtering by year, venue, etc., as seen in VisPubs (\cite{lange_vispubs.com}).
However, it is not the only way --- VITALITY \cite{narechania_vitality_2022} uses transformers to place similar papers near each other in an embedding to ``promote serendipitous discovery of relevant literature.'' \footnote{We noticed this paper while play-testing Colon. So this is officially the first citation that results from a joyful discovery while playing one of the VisPubs Games!}
Although both of these approaches have merit, our games take an even more straightforward approach --- they simply select papers and authors randomly from the VisPubs dataset \cite{lange_vispubs.com}.
This approach of randomly serving papers is not unfounded. For instance, the IEEE VIS conference website takes a similar approach to sorting papers on the website, ``by serendipity'' (See Figure~\ref{fig:vis_shuffle_serendipity}).

The \textbf{visualization research community} is a small group of academics with a highly specific set of common interests; however, we are physically distributed throughout the world. We made these games for visualization researchers, and we believe the community will enjoy them.
Although others can play the game, we are skeptical that anyone outside this community will find these games particularly fun.
Most research claims that the applicability of the work to a broad audience strengthens the contribution. In our case, we argue the opposite. The extreme specificity of the intended audience could serve to bring that community together.
Furthermore, the ease of playing the game online and sharing results lends itself to the online communities and discussions that our physically distributed nature requires.

Most researchers do not enjoy \textbf{self promotion}, but many find that it is beneficial or even necessary. One of the two games, Authored, is centered around visualization researchers. This aspect of the game provides passive exposure for researchers that requires no effort on the researchers' part. Additionally, it can support lower effort ways to promote yourself or other researchers online --- ``I was on Authored, yay!'' or ``Professor X was on Authored, I always love their insight on visualization design.''

Last but not least, these games should be fun. To reach these final two game designs, we explored dozens of ideas, many of which were technically feasible and would've supported the same conceptual arguments as the previous points, but they probably would not have been fun. We felt that many game ideas were actually testing how well people have the titles and authors of the roughly 5,000 papers on VisPubs memorized. This is a memory test, not a fun game. Furthermore, it would be prohibitively difficult for novice researchers. The games we designed are playable without a photographic memory of every visualization publication. Players with extensive knowledge of visualization research will have an advantage, but novices can still succeed.
We cannot guarantee that every visualization researcher will enjoy these games, but we hope many will. If nothing else, we enjoy playing them.

\section{Game Designs}
Why are you reading how a game works? Just play it! Do it! \href{https://games.vispubs.com}{games.vispubs.com}.
We also include a brief description of both games here for the nerds who like to read.

\subsection{Colon: Catchy Name First, Meaning Second}
Many academic papers include a colon in the title. There are several patterns for this. One common approach is to include a catchy name for the tool or technique before the colon and the explanation of that name in the second half. Sometimes the connection between the two parts is obvious, sometimes it is more oblique, and other times there is no explicit connection.
The structure of visualization papers has been explored at alt.vis already by Lang\footnote{Not to be confused with Lange et al.\ on this paper!} et al.\ \cite{lang_name_2022}, who generate visualization papers automatically.
We exploit this structural variety and turn it into a game --- ``Colon: Catchy Name First, Meaning Second'' (See Figure~\ref{fig:teaser}a). The goal of the game is to reconnect four papers that have been split at the colon and shuffled.

\subsection{Authored}
Authored is heavily inspired by the New York Times game, Connections. In Connections, players identify groups of four words based on abstract connections. In our game, Authored, we present a grid of nine papers, and the player must locate three groups of three papers (See Figure~\ref{fig:teaser}b). Each group contains a common author who is on all three papers. If the player guesses incorrectly, additional information is shown in the game as hints, such as the publishing venue and year. The papers are selected by first randomly choosing three authors with at least three publications in VisPubs \cite{lange_vispubs.com}. This approach ensures that the game isn't biased towards frequently selecting highly prolific authors. 

\section{Why VisPubs Games Are Awesome}

In case you need further convincing to play our games, this section is for you. If not, head over to \href{https://games.vispubs.com}{games.vispubs.com}.

\subsection{They Are Not Educational (a.k.a. they are fun)}

VisPubs Games games are not edutainment. This is, frankly, their greatest strength. They do not pretend that layering a worksheet beneath a fun veneer constitutes fun. They do not require players to learn a lesson before they are allowed to have a good time. In fact, we are proud to say that VisPubs Games are aggressively disinterested in being educational. Any learning that occurs is a side effect --- a fortunate accident, like getting your daily fiber from eating a really good burrito.

These games are not chocolate-covered broccoli~\cite{laurel_utopian_2001}. They are chocolate. And yes, perhaps it happens to be 85\% dark and rich in antioxidants, but that is not why you are here. You are here because it is delicious. We designed games that are actually fun. If they also spark insights about trends in visualization research, authors, or paper titles, well, that is a happy bonus. But to be clear: we are not trying to sneak in vitamins.

\subsection{They Scale (a.k.a. you will never run out of fun)}

Moreover, VisPubs Games games are also highly scalable --- you're unlikely to ever run out of fun. For instance, the current VisPubs~\cite{lange_vispubs.com} database contains 2,202 papers with a colon in the title. That is already a rich design space: there are
\[
\binom{2202}{4} = 978,\!341,\!827,\!999
\]
possible unique combinations of four-paper Colon-themed games. To estimate how many games one might need to play to see them all, we can plug into the classic \textit{coupon collector's problem}, yielding:
\[
\frac{2202 \cdot \ln(2202)}{4} \approx 4,\!237
\]
games, on average, to see each Colon paper appear at least once.

Similarly, we are not going to run out of Authored games anytime soon either. There are 1,741 authors with three or more papers, which means, on average, you need to play about as many Authored games to have seen all authors:
\[
\frac{1741 \cdot \ln(1741)}{3} \approx 4,\!331
\]
The numbers escalate quickly, offering a virtually inexhaustible well of game content, remixable for novelty, chaos, and discovery.
And that is just the beginning. Since VisPubs Games links directly to VisPubs \cite{lange_vispubs.com}, as new papers are added, the number of available games will grow automatically and exponentially.

\section{Authors' Reflections}

We have reflected on our experience while playing Colon and Authored. Firstly, they are fun. It could be the bias of a creator playing their own game, but we all find these games to be an enjoyable experience.
In addition, we have found that playing the game has led to joyful discoveries of interesting and serendipitously relevant papers, such as VITALITY \cite{narechania_vitality_2022}.
Between Authored and Colon, there is a nice variety of difficulty. Colon is a much easier game and can be completed very quickly. Alternatively, Authored can be difficult and more time-consuming. This variety is beneficial for players of different experience levels and varying time availability.
Searching for patterns to solve the game does result in more engagement with a paper title than simply reading or skimming the paper from a list.
While playing Authored, the most common strategy we employed was to look for the same sub-sub-sub-field across papers; however, we also encountered situations where we discovered a researcher we were familiar with conducted research in an area we weren't expecting. In addition to matching research areas, we also found that in some games the language style was a significant clue in both Colon and Authored.

\section{Conclusion and Future Work}
In conclusion, VisPubs Games is an entertaining and novel way for visualization researchers to engage with academic literature. These games can facilitate joyful discovery of research and researchers and serve as an artifact our community can use to interact with each other and strengthen our ties. In the future --- if inspiration strikes --- we may develop additional games.

\acknowledgments{
Thank you to everyone in the visualization community who listened to this silly idea, provided game design suggestions, and played early versions of the game.
}

\bibliographystyle{abbrv-doi}

\bibliography{ref}
\end{document}